\documentclass[aps,prl,twocolumn,superscriptaddress,
,floatfix]{revtex4}
\usepackage{graphicx}
\usepackage{amsmath,amssymb}
\usepackage{color}

\begin{document}

\title{Losses resistant verification of quantum non-Gaussian photon statistics}

\author{Riccardo Checchinato}
\affiliation{Department of Physics, Stockholm University, 10691 Stockholm, Sweden}

\author{Jan-Heinrich Littmann}
\affiliation{Department of Physics, Stockholm University, 10691 Stockholm, Sweden}

\author{Luk\'a\v{s} Lachman}
\affiliation{Department of Optics, Palack\' y University, 17. listopadu 12, 77146 Olomouc, Czech Republic}

\author{Jaewon Lee}
\affiliation{Department of Physics, Stockholm University, 10691 Stockholm, Sweden}

\author{Sven H\"{o}fling}
\affiliation{Technische Physik, Physikalisches Institut and W\"urzburg-Dresden Cluster of Excellence ct.qmat, Universit\"at W\"urzburg, Am Hubland, D-97074 W\"urzburg, Germany}

\author{Christian Schneider}
\affiliation{Institut of Physics, University of Oldenburg, D-26129 Oldenburg, Germany}

\author{Radim Filip}
\affiliation{Department of Optics, Palack\' y University, 17. listopadu 12, 77146 Olomouc, Czech Republic}

\author{Ana Predojevi\'{c}}
\email{ana.predojevic@fysik.su.se}
\affiliation{Department of Physics, Stockholm University, 10691 Stockholm, Sweden}

\begin{abstract}
Quantum non-Gaussian states of light have fundamental properties that are essential for a multitude of applications in quantum technology. However, many of these features are difficult to detect using standard criteria due to optical losses and detector inefficiency. As the statistics of light are unknown, the loss correction on the data is unreliable, despite the fact that the losses can be precisely measured. To address this issue, we employ a loss-mitigated verification technique utilising quantum non-Gaussian witnesses, which incorporate the known optical losses and detector inefficiency into their derivation. This approach allows us to address the considerable challenge of experimentally demonstrating unheralded quantum non-Gaussian states of single photons and photon pairs.
\end{abstract}

\maketitle

The development of quantum technology is contingent upon the utilisation of quantum non-Gaussian states \cite{Walschaers2021, Lachman2022, Rakhubovsky}, which are indispensable for sensing \cite{Wolf2019, McCormick2019} and hybrid systems comprising quantum bits \cite{Ofek2016, Ni2023, Sivak2023}. The advancement of quantum systems for photonic sources in atomic \cite{Higginbottom, Daiss, Hacker, Magro}, solid-state \cite{gauss1, StrakaPRL}, and nonlinear optics \cite{Ourjoumtsev2006, Neergaard2006} platforms is hindered by the difficulty in detecting quantum non-Gaussian aspects of photons. A significant challenge for many new sources and their initial tests is the loss of light during transfer from the source to the measurement and the inefficiency of the detection process. Such loss, exceeding three decibels, limits the direct observation of a witness of quantum non-Gaussian light using a negative Wigner function \cite{Lee1991,Lee1992}.

A significant advance was made by quantum non-Gaussianity witnesses, enabling the three-decibel limit to be surpassed and quantum non-Gaussianity to be identified directly \cite{Filip2011, JezekPRL, Genoni2013, Hughes2014, Park2017, Happ2018}. Subsequently, this methodology has been successfully deployed in nonlinear optics \cite{Straka2019, Baune2014}, atomic physics \cite{Higginbottom, Mika2022}, and solid-state sources \cite{gauss1, StrakaPRL}. The optical loss and detection inefficiency can be accurately quantified and characterized, which is advantageous. Nevertheless, the direct correction of the unknown quantum statistics of light for this loss is typically unreliable. Therefore, we incorporate the precisely characterized inefficiency into the construction of the quantum non-Gaussianity criteria. This allows conclusive proof of unheralded quantum non-Gaussian single-photon states and their coincidences generated from quantum emitters.

The efficiency of optical transmission, light collection, and detection is limited for many photonic measurements widely used in quantum physics. Failure to achieve unity efficiency can be modelled as the presence of a lossy channel with transmission $\eta$ \cite{MandelBook}. We derive a criterion for certifying quantum non-Gaussianity that incorporates this model of non-ideal efficiency and experimentally verify it. Quantum non-Gaussianity can be observed in a variety of states including single photons and pairs of photons \cite{JezekPRL, Mika2022}. To address single photons, we consider a measurement that employs a beamsplitter and two non-ideal detectors, $D_1$ and $D_2$. We define the probability $P_{1,\eta}$ of a click on detector D$_1$ and the probability $P_{2,\eta}$ of a coincidence click on both detectors D$_1$ and D$_2$ per generated state. Both probabilities $P_{1,\eta}$ and $P_{2,\eta}$ depend on the transmission of the lossy channel $\eta$. While there are criteria \cite{JezekPRL, LachmanPRL} for quantum non-Gaussianity that include all mixtures of Gaussian states, the states studied experimentally may not meet this criterion due to high optical losses, which result in low overall detection efficiency. In a regime of photonic states with $P_{2,\eta}\ll P_{1,\eta}$, we can approximate the criterion derived in \cite{JezekPRL} with $P_{1,\eta}^3 \gtrsim P_{2,\eta}/4$ \cite{StrakaPRL}. 

Here we present a novel approach that takes into account measurement losses. To this end, we have derived the threshold by maximizing the linear combination $F_{\alpha,\eta} \equiv P_{1,\eta}+\alpha P_{2,\eta}$ over mixtures of pure Gaussian states, where we consider that the probabilities $P_{1,\eta}$ and $P_{2,\eta}$ were detected employing a lossy channel (detector) with a given efficiency $\eta$. 
 
Additional information on the numerical derivation of the threshold can be found in \cite{supp}. In a regime of states with low probability $P_{2,\eta}$, the criterion takes on the approximate form $P_{1,\eta}^3 \gtrsim \eta/ \left[4 (2-\eta)\right]P_{2,\eta}$. This inequality indicates that the criterion is more lenient than the one given in \cite{StrakaPRL}.

\begin{figure*}[t]
    \centering
    \includegraphics{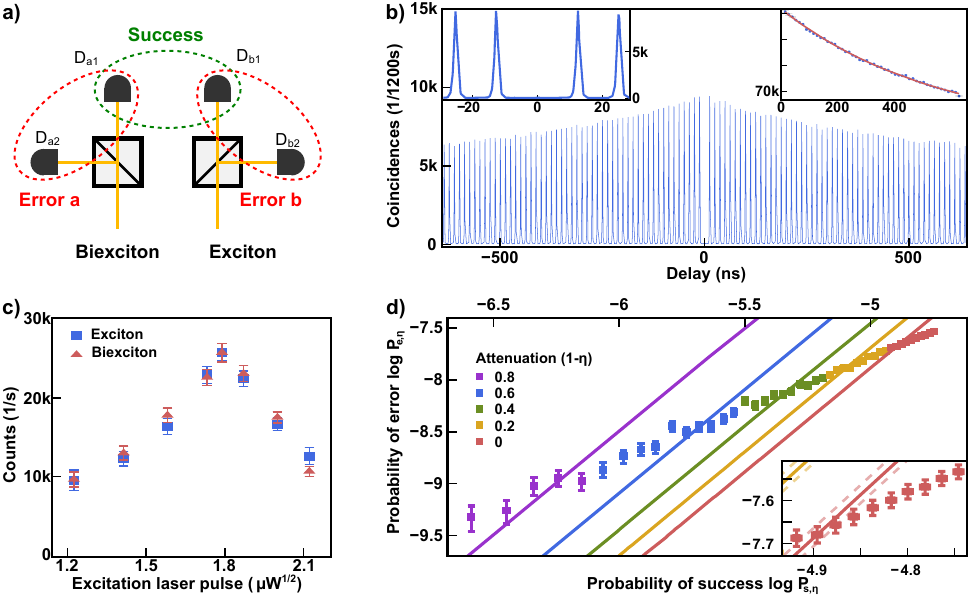}
    \caption{\label{fig:1} (a) Schematic of the measurement setup. Coincidences between D$_{1a}$ and D$_{2a}$ (indicated by the green circle) are considered to be successful outcomes while coincidences between either D$_{1a}$ and D$_{1b}$ or D$_{2a}$ and D$_{2b}$ (indicated by the red circles) are considered to be error events. (b) Autocorrelation function measured using exciton photons. The inset in the top left-hand corner of the figure shows the autocorrelation function near zero delay. The measurement gives the value of $g^{(2)}(0)= 0.0154(4)$. The inset on the top right shows the summed area under each peak. These values were fitted with an exponential function to derive the blinking rate. (c) Result of the Ramsey interference measurement. We determined the excitation laser power corresponding to a $\pi/2$ pulse by estimating where the Ramsey fringes reach their maximum amplitude. (d) Comparison between our data and the non-Gaussian coincidences criterion. Each line represents a threshold of the criterion. The color coding of the lines indicates the attenuation introduced. The point with the highest probability of success (also shown in the inset) has the coordinates corresponding to the experimentally determined $P_{e, \eta}$ and $P_{s,\eta}$. We find this point to be 11.52 standard deviations from the criterion threshold. The remaining points were derived by emulating the attenuation of the optical signal (loss) by undersampling. The attenuation was varied in the same range as for the criterion thresholds and in steps of 0.02. The color of the points and lines (thresholds) changes when the introduced attenuation exceeds 0.2. The points positioned to the right of the corresponding threshold indicate the respective quantum state is non-Gaussian. Inset: Area of the plot with the lowest attenuation in greater detail. The dashed lines represent the uncertainty associated with threshold, which arises from the measurement error of the setup transmission and the imbalance of the beamsplitters.}
\end{figure*}

To certify the non-Gaussianity of photon pairs \cite{LachmanPRL}, we consider that each photon in the pair occupies a different mode and assume that both of them propagate through the same lossy channel. To perform the measurement, each mode is sent to a beamsplitter followed by two detectors D$_{m1}$ and D$_{m2}$ where $m\in\{a,b\}$ represents the mode index, as shown in Fig.\ref{fig:1}a. For a given transmission $\eta$ of the lossy channel, we introduce the probability $P_{s,\eta}$ of a coincidence between detectors D$_{a1}$ and D$_{b1}$. Similarly, we define the probability $P_{e,m,\eta}$ of a coincidence between detectors D$_{m1}$ and D$_{m2}$ pertaining to the same mode $m$. Based on the probabilities $P_{s,\eta}$ and $P_{e,\eta}\equiv \left[P_{e,a,\eta}+P_{e,b,\eta}\right]/2$ we can, in principle, derive a criterion capable of rejecting a generic multi-mode Gaussian state. In practice, however, the derivation of such a criterion is computationally demanding, even in the case of ideal detection \cite{LachmanPRL}. This is due to necessity of maximizing over the large number of parameters required to identify a generic multi-mode Gaussian state.

As this is a pervasive issue in quantum state benchmarking, we put forth a viable solution: the derivation of more specific, pertinent, and indispensable models for the source under examination. Therefore, we reduce our study to the subset of multi-mode states $\rho_N=\Pi_{i=1}^{N} \otimes \rho_i(\lambda_i)$ with two-mode Gaussian state $\rho_i(\lambda_i)$ defined as:
\begin{equation}\label{SPDC}
    \rho_i(\lambda_i)=\frac{1}{1+\lambda_i^2}\sum_{n=0}^{\infty}\lambda_i^n |n\rangle_{a_i} \langle n|\otimes |n\rangle_{b_i} \langle n|,
\end{equation}
where $\lambda_i$ determines the distribution of photons correlated in the $a_i$-th mode and the $b_i$-th mode. The state $\rho_N$ represents a model state of light emitted from spontaneous parametric down-conversion process \cite{LachmanPRL,SPDC}. After considering the potential impact of a lossy channel on the state $\rho_N$, we express the measured probabilities $P_{s,\eta}$ and $P_{e,\eta}$ as a function of the parameters $\lambda_i$ in the state $\rho_N=\Pi_{i=1}^{N} \otimes \rho_i(\lambda_i)$. In order to establish a criterion for $N$ modes, we need to maximize the linear combination $F_{\alpha}(\rho_N)=P_{s,\eta}+\alpha P_{e,\eta}$ over $N$ parameters $\lambda_1,...,\lambda_N$. We can identify for each $N$ a threshold in terms of $P_{s,\eta}$ and $P_{e,\eta}$ that covers all states $\rho_N$. Although the maximizing task becomes challenging for large $N$, we can solve it for an arbitrary $N$ by considering an experimentally relevant regime of small $P_{e,\eta}$, as summarized in \cite{supp}. In this limit the criterion follows approximately as:
\begin{equation}
    P_{s,\eta}> \eta \frac{N}{2\sqrt{N(N+1)}}\sqrt{P_{e,\eta}}.
    \label{thres}
\end{equation}

In order to encompass all possible scenarios, we consider the asymptotic multi-mode limit of $N \gg 1$, where the left side attains its maximum over $N$. This provides the most rigorous version of the criterion. This enables us to reject the multimode two-mode squeezed Gaussian states $\rho_N$ for any number of equally occupied independent modes modes.

Our photon-pair source consists of a quantum dot embedded in a micropillar cavity with a diameter of 2.74$\mu m$ \cite{lowQ}. During the measurements the quantum dot was kept in a closed-cycle cryostat at a temperature of 4.856(8)\,K. We excited the quantum dot resonantly using a pulsed Ti:Sapphire laser with a repetition rate of 80\,MHz. To ensure the generation of pairs of photons the quantum dot was driven using two-photon resonant excitation of the biexciton \cite{twophoton}. The excess laser scattering was eliminated by employing a combination of spectral and polarization filtering. The emitted biexciton-exciton photon pairs were spectrally separated using a diffraction grating and coupled in respective single-mode fibres. The biexciton and exciton photons were detected using avalanche photo-diode (APD) detectors.

The measurement scheme used to assess our source against the non-Gaussian coincidences criterion is depicted in Fig.~\ref{fig:1}a. The fiber-coupled biexciton (exciton) photons were transmitted through a fibre beamsplitter and subsequently detected by an APD detector. We employed a total of four APD detectors, two for each fiber beamsplitter we used.
The coincidences recorded by detectors D$_{\text{a1}}$ and D$_{\text{a2}}$ (D$_{\text{b1}}$ and D$_{\text{b2}}$) are designated as error events, as they indicate a simultaneous emission of two biexciton (exciton) photons. Conversely, the coincidences recorded between the detectors D$_{\text{a1}}$ and D$_{\text{b1}}$ are considered to be successful events.
The probability of success is defined as $P_{s,\eta} = C_{s}/C_{0}$. Here, $C_{0}$ is the photon pair generation rate, while $C_{s}$ is the rate of coincidences recorded between D$_{\text{a1}}$ and D$_{\text{b1}}$.
The probability of an error event $P_{e, \eta}$ is defined as $(C_{e,a}+C_{e,b})/2C_{0}$. Here, $C_{e,a}$ ($C_{e,b}$) is the rate of coincidences recorded by detectors D$_{\text{a1}}$ and  D$_{\text{a2}}$ (D$_{\text{b1}}$ and  D$_{\text{b2}}$).

To ascertain the values of $P_{s,\eta}$ and $P_{e,\eta}$ we require to accurately determine $C_{0}$. The photon pair generation rate for a resonantly excited single quantum dot emitter is dependent on the emission probability and the blinking rate. The blinking rate is commonly estimated using an autocorrelation measurement (shown in Fig. ~\ref{fig:1}b), where it induces an exponential decay of the height of the peaks \cite{blinking}. 
By fitting the exponential decay, we estimated that the effect of blinking reduces the rate at which photon pairs are generated to 56.6(6) $\%$ of the laser repetition rate. The emission probability is a function of the pulse area of the excitation laser and is commonly determined by Rabi oscillations measurements. However, the accurate fitting of the Rabi oscillation demands coherence to be preserved even for very high pulse area.
We resourced to a simpler approach where we employed Ramsey interference. Namely, the amplitude of the Ramsey interference fringes increases with the pulse area, reaching maximum at $\pi/2$ pulse, to gradually fall to zero at $\pi$ pulse (shown in Fig. ~\ref{fig:1}c). Consequently, Ramsey interference can be used to determine the laser power that equals the excitation pulse area of $\pi/2$.

We performed the measurement using an excitation laser with repetition rate of 80\,MHz and a pulse area of $\pi/2$. These two parameters combined with the above-determined blinking yield a photon pair generation rate of 22.64(25)\,MHz. However, due to suppression of the laser excess scattering by means of polarization only the photon pairs with vertical polarization will take part in the measurement. Therefore, we can consider that during the measurements we performed the photon pair generation rate was equal to $C_{0}$=11.32(12)\,MHz.

The coincidence counts were sampled within a 1408\,ps coincidence window, selected to facilitate the efficient detection of both biexciton and exciton photons with lifetimes of 249.8(7)\,ps and 397.2(9)\,ps, respectively. Our measurements yielded 244113(494) coincidence counts recorded between D$_{\text{a1}}$ and D$_{\text{b1}}$ in a measurement time of 1200\,s. During the same time span we recorded 365(18) coincidence counts between D$_{\text{a1}}$ and  D$_{\text{a2}}$ and 430(21) between D$_{\text{b1}}$ and  D$_{\text{b2}}$. These number yield $P_{s, \eta}=1.797\pm0.020\cdot 10^{-5}$ and $P_{e, \eta}=2.93\pm0.11\cdot 10^{-8}$.

The criterion for non-Gaussian coincidences accounts for losses of the setup and the detection. The overall losses of the optical elements were quantified by utilising a laser with a wavelength corresponding to that of the emitted photons. The measurement yielded an optical transmission of 0.534(16). The fibre coupling efficiency was determined to be 0.692(2). Due to the fibre mating losses, only 0.7948(4) of the light coupled into the single-mode fibres was fed into fibre beamsplitters. The fibre coupling and mating losses were averaged over biexciton and exciton. It is assumed that the APD quantum efficiency is 0.5 and is the same for all the APD detectors used in the measurement. This value is significantly higher than the actual quantum efficiency, as stated in the specifications. It can thus be estimated that the total losses of the apparatus, including the detectors, are at least 0.8533(34), indicating an $\eta$ of at most 0.1467(34).

\begin{figure}[t]
    \centering
    \includegraphics[width=0.95\linewidth]{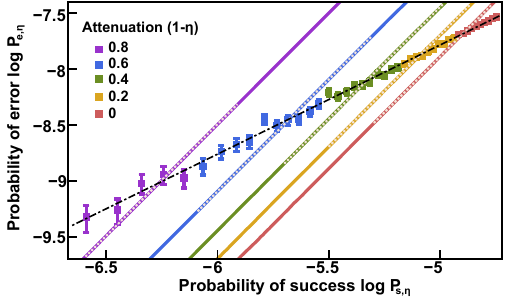}
    \caption{\label{fig:2} Calculation of the non-Gaussian depth by data fitting. The different criterion thresholds, represented by solid coloured lines, are second-order expansions of the criterion in the vicinity of the data points. A linear fit is performed on each of the criterion thresholds and on the complete data set. Subsequently, the non-Gaussian depth of the states was obtained from the intersection between the criterion threshold and the data fit. The results are given below in Table \ref{tab:result}.}
\end{figure}

\begin{table}[h!]
\caption{The non-Gaussian depth of the state determined for different levels of attenuation.}
\label{tab:result}
\begin{ruledtabular}
\begin{tabular}{cccc@{}r@{\quad}}
\multicolumn{1}{c}{Attenuation $(1-\eta)$}&
\multicolumn{1}{c} {non-Gaussian depth (dB)}   \\ \hline
$0$  &    $\hspace{1.4 mm}0.764 \pm 0.039$  \\
$0.2$  &  $\hspace{1.4 mm}0.740  \pm 0.035 $   \\
$0.4$ &   $\hspace{1.4 mm}0.710 \pm 0.033 $   \\
$0.6$ &   $\hspace{1.4 mm}0.667  \pm 0.035$    \\
$0.8$ &  $\hspace{1.4 mm}0.595 \pm 0.049$   \\
\end{tabular}
\end{ruledtabular}
\end{table}

The experimentally determined probability of success and error (illustrated in Fig.~\ref{fig:1}d as the point with the highest success probability) substantiates the assertion that our source emits non-Gaussian coincidences.
The obtained result is 11.52 standard deviations away from the criterion threshold.
Moreover, the effect of optical losses in the range from zero to 0.8 was emulated by means of undersampling of the data. Figure Fig.~\ref{fig:1}d illustrates the impact of attenuation on the probability of error and success. 
This approach allows for the emulation of optical losses in the channel up to the level of the detector dark counts. This is so, as coincidences between dark counts are removed by undersampling, whereas they would remain unaltered by the introduction of losses in the optical path. However, we estimated that such events would contribute less than 0.031 coincidences on average over the course of the entire measurement. It can thus be concluded with a high degree of confidence that the results presented accurately reflect the process of experimentally introducing optical losses.  

To test the robustness of the criterion in the presence of losses, we proceeded to estimate the non-Gaussian depth \cite{StrakaPRL} of our state. In order to achieve this, we derived an approximate analytical form for the criterion. The complete range of data points was also fitted, as illustrated in Fig.~\ref{fig:2}. Subsequently, the point of intersection between the each threshold and the fit was calculated. The respective non-Gaussianity depths are shown in Tab.~\ref{tab:result}.

\begin{figure}[t]
    \centering
    \includegraphics[width=0.95\linewidth]{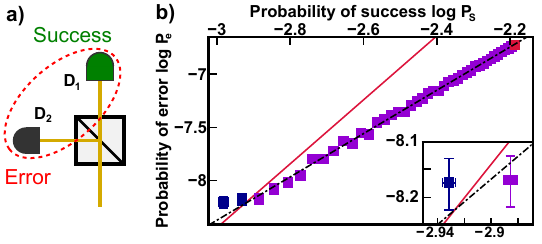}
    \caption{\label{fig:3} (a) Experimental scheme to certify quantum non-Gaussianity of unheralded single photons states. Coincidence between detectors on D$_\text{1}$ and D$_\text{2}$ are denoted as error events while a click on detector D$_\text{1}$ is considered to be the success event. (b) Comparison between our data and the criterion for quantum non-Gaussianity of single photon states \cite{supp}. The solid red line indicates the criterion. It marks the transition between Gaussian states (blue squares) and non-Gaussian states (purple squares). For the criterion threshold plotted here, the losses of our measurement apparatus were not taken into account; nevertheless, our data show that the detected state is quantum non-Gaussian and positioned 183 standard deviations from the criterion threshold. Analogous to the data in Fig.~\ref{fig:1}(d), we have undersampled to simulate optical losses in steps of $0.02$.  The obtained data has been fitted (dashed black line) to determine the non-Gaussian depth ($7.41 \pm 0.11$ dB) . The inset shows the states that cross the threshold. }
\end{figure}

In addition to the results shown so far, we have also used the same source to certify quantum non-Gaussianity for unheralded single photon states. To achieve this, we deliberately excluded the biexciton data from our analysis and focused only on the exciton photon emission from our device. The experimental implementation in this case involves a stream of exciton single photons and a beamsplitter, as shown in Fig. \ref{fig:3}a.

The results are shown in Fig. \ref{fig:3}b. We have used single photon non-Gaussianity criterion \cite{supp}, which defines the probability of error $P_e$ using the probability of success $P_S$ and the transmission of a beamsplitter $T_{bs}$ (0.5166(3)) in front of the two detectors $D_1$ and $D_2$: 
\begin{equation}
P_e < 2\cdot P_S^3 \cdot \frac{1 - T_{bs}}{T_{bs}^2}.    
\end{equation}
The measured state is 183 standard deviations from the criterion threshold. We determined a non-Gaussian depth of $7.41 \pm 0.11$ dB  via a fit to the data points, which corresponds to an $\eta$ of 18.17(21)$\%$. 

In conclusion, we have developed two new criteria for the quantum non-Gaussianity of single photon and photon pair states. Both of our criteria can handle known optical losses, making them a very powerful tool for studying states of light in lossy and long-distance channels. We have tested these criteria in practice using a quantum dot-based source of photon pairs.

The proposed methodology can be further extended to the detection of multiphoton quantum non-Gaussianity \cite{Straka2018} and also to the loss-resistant form of a hierarchy of quantum non-Gaussian states \cite{Straka2019}, allowing the identification of increasing quantum non-Gaussian aspects when the stellar hierarchy \cite{Chabaud2020, Chabaud2021} or even more general approaches \cite{Straka2019} remains challenging in the experiment.

\begin{acknowledgments}

A.P. would like to acknowledge the Swedish Research Council (grant 2021-04494). J. L. was supported by the Knut \& Alice Wallenberg Foundation (through the Wallenberg Centre for Quantum Technology (WACQT)). We acknowledge financial support by the State of Bavaria. L.L. acknowledges the grant 23-06015O and R.F. received support from the grant 21-13265X, both provided by the Czech Science Foundation.
\end{acknowledgments}

\end{document}